\documentclass[epj]{svjour}
\usepackage{graphics}
\begin{document}
\title{High precision half-life measurement of $^{95}$Ru, $^{95}$Tc and $^{95m}$Tc with $\gamma$-spectroscopy}
\author{T. N. Szegedi\inst{1,2} \and \'A. T\'oth\inst{1} \and G. G. Kiss\inst{2} \thanks{\emph{corresponding author:} ggkiss@atomki.mta.hu} \and Gy. Gy\"urky\inst{2} 
%
}                     
%
%
\institute{University of Debrecen, Debrecen, H-4032, Hungary \and Institute of Nuclear Research, Debrecen, H-4032, Hungary}
\date{Received: date / Revised version: date}
%
\abstract{
The precise knowledge of the half-life of the reaction product is of crucial importance for a nuclear reaction cross section measurement carried out with the activation technique. The cross section of the $^{92}$Mo($\alpha$,n)$^{95}$Ru reaction was measured recently using this experimental approach. The preliminary results indicated that the literature half-life of $^{95}$Ru, derived about half a century ago, is overestimated. Therefore, the half-lives of $^{95}$Ru and its daughter isotope $^{95}$Tc and $^{95m}$Tc have been measured with high precision using $\gamma$-spectroscopy. The results are t$_{1/2}$\,=\,1.6033\,$\pm$\,0.0044\,h for $^{95}$Ru, t$_{1/2}$\,=\,19.258\,$\pm$\,0.026\,h for $^{95}$Tc and t$_{1/2}$\,=\,61.96\,$\pm$\,0.24\,d for $^{95m}$Tc. The precision of the half-life values has been increased, consequently the recently measured $^{92}$Mo($\alpha$,n)$^{95}$Ru activation cross section will become more precise.
} 
\maketitle
\section{Introduction}
\label{int}

The bulk of the isotopes heavier than iron are synthesized via neutron capture reactions in the so-called s and r processes \cite{s_review,r_review}. However, on the proton-rich side of the valley of stability there are -- depending on the astrophysical model -- 30-35 mostly even-even proton-rich species. These, so-called, p-nuclei cannot be synthesized by neutron capture reactions, since they are separated by unstable short-lived nuclei from the path of both the s and r processes \cite{p_review}. According to our knowledge, in the production of these isotopes, ($\gamma$,n), ($\gamma$,p) and ($\gamma$,$\alpha$) photodisintegration reactions play key role and the process takes place either in type IA supernovae or in Type II (core collapse) supernovae \cite{p_review,tra11}. 

\begin{figure}
\resizebox{0.5\textwidth}{!}{\includegraphics{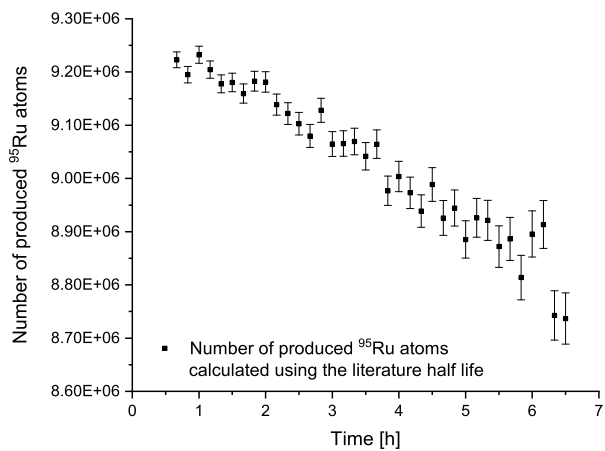}}
\caption{The number of $^{95}$Ru isotopes produced in the $^{92}$Mo($\alpha$,n)$^{95}$Ru reaction as a function of time after the activation calculated from the 336.40 keV peak area using the literature half-life (t$_{1/2}$\,=\,1.643\,$\pm$\,0.013\,h). The slope of the data indicates that the half-life value may be overestimated.}
\label{fig:old_half}       
\end{figure}

The astrophysical modeling of this nucleosynthesis scenario requires an extended reaction network calculation, the necessary cross sections are taken from the statistical model. The predictions of the statistical model can be tested and its input parameters -- such as optical model potentials, level densities and $\gamma$-ray strength functions -- can be optimized by measuring the cross sections of charged-particle induced reactions, the inverses of the relevant photodisintegrations. Such cross section measurements are often carried out using the activation technique \cite{gyu19}. The application of this experimental approach requires precise information on the decay properties of the reaction product(s). Accordingly, the half-life (t$_{1/2}$) of the resulted isotope and the intensities of the $\gamma$-rays emitted after the $\beta$-decay have to be known to derive the number of created reaction product nuclei.

\begin{figure*}
\resizebox{1.0\textwidth}{!}{\includegraphics{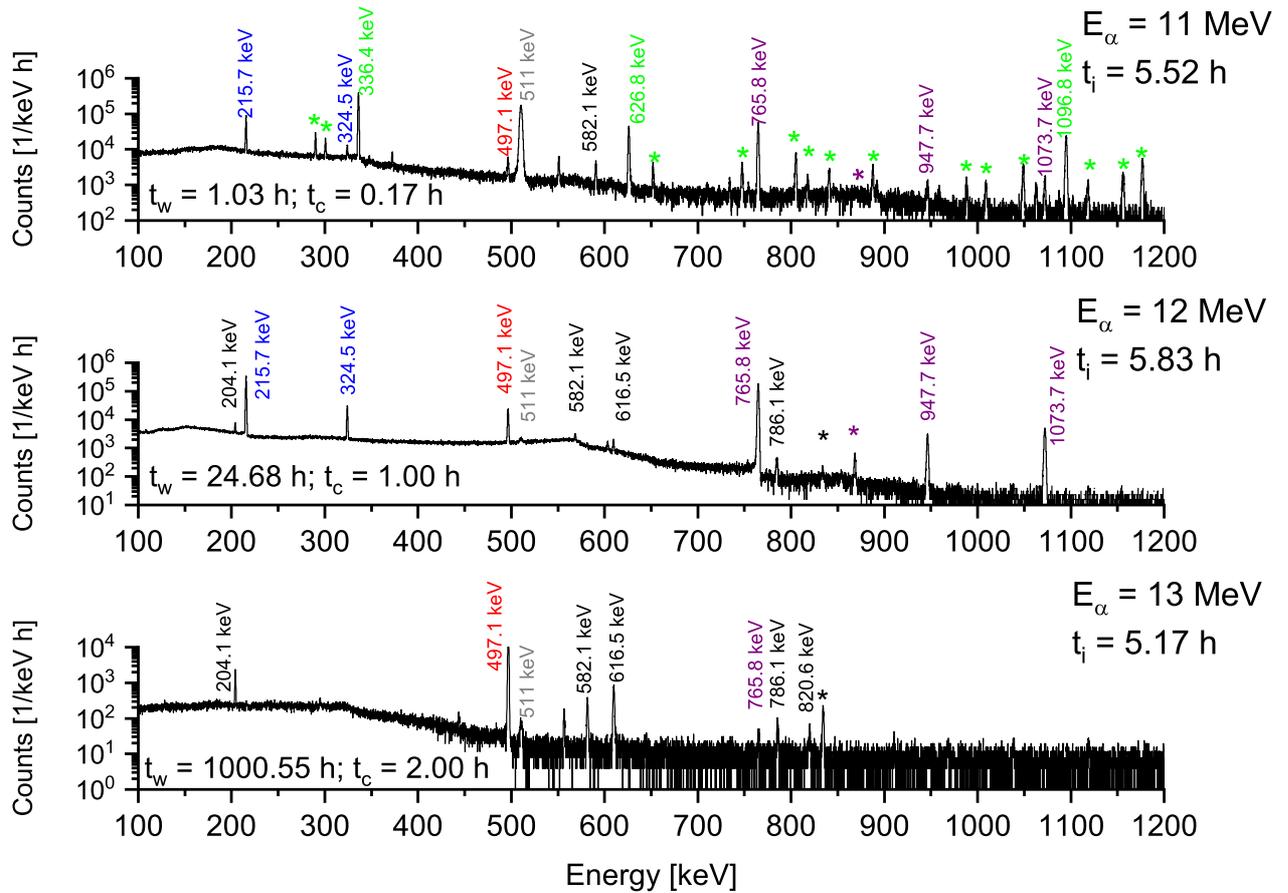}}
\caption{A t$_c$\,=\,0.17\,h (A), t$_c$\,=\,1\,h (B) and t$_c$\,=\,2\,h (C) $\gamma$-spectrum measured t$_w$\,=\,1.03\,h (A), t$_w$\,=\,24.68\,h (B) and t$_w$\,=\,1000.55\,h (C) after the end of the irradiation by E$_{\alpha}$\,=\,11.0\,MeV (A), E$_{\alpha}$\,=\,12.0\,MeV (B) and E$_{\alpha}$\,=\,13.0\,MeV (C) $\alpha$ beams. The peaks used to derived the half-lives of $^{95}$Ru, $^{95}$Tc and $^{95m}$Tc are indicated with green, purple and black numbers, respectively (weaker transitions, not used for the t$_{1/2}$ determination are indicated by asterisks). $\gamma$-lines originated from the $\beta$-decay of $^{97}$Ru and $^{103}$Ru were used to monitor the stability of the $\gamma$-counting setup, the corresponding peaks are labeled with blue and red, respectively.}
\label{spectra}       
\end{figure*}

To constrain the parameters of the $\alpha$-nucleus optical potential, a new measurement for the $^{92}$Mo($\alpha$,n)$^{95}$Ru reaction cross section is in progress at the Institute of Nuclear Research (Atomki). The number of reaction products is deduced from the yield of the well-known $\gamma$-radiation \cite{she07} emitted after the $\beta$-decay of $^{95}$Ru. From the measured activities, the initial number of the produced $^{95}$Ru nuclei has been calculated using the adopted half-life value (t$_{1/2}$\,=\,1.643\,$\pm$\,0.013\,h \cite{NDS}; derived as the weighted mean of the results of two experiments: t$_{1/2}$\,=\,1.650\,$\pm$\,0.017\,h \cite{pin68}, t$_{1/2}$\,=\,1.632\,$\pm$\,0.021\,h \cite{por70} carried out about half a century ago). The preliminary results, shown in Fig. \ref{fig:old_half}, clearly show a decreasing tendency which may indicate that the half-life of the $^{95}$Ru is overestimated. 

Furthermore, the cross section of the investigated reaction can also be derived by measuring the activity of the daughter isotope of the reaction product. However, the
adopted half-life value of $^{95}$Tc is based on only one measurement \cite{NDS,NDS2}\footnote{Two further experiments, carried out in 1948, aimed at the determination of the half-life of the ground state of $^{95}$Tc \cite{Mot48,Egg48}. However, these data (t$_{1/2}\,=\,$20\,$\pm$\,2\,h and t$_{1/2}$\,=\,20.0\,$\pm$\,0.5\,h) were not taken into account when the adopted half-life value of $^{95}$Tc was derived \cite{NDS2} probably due to the fact that in 1948 the metastable state was not identified yet and therefore the data of \cite{Mot48,Egg48} is not corrected for its IT decay.} carried out almost 50 years ago using a NaI(Tl) detector \cite{vin62}. In table 1 of \cite{vin62} the $\gamma$-transitions assigned to the $\beta$-decay of $^{95}$Tc are listed, however, according to \cite{she07,NDS} some of the transitions belong also to the decay of the metastable state of $^{95}$Tc ($^{95m}$Tc) or $\gamma$-transitions emitted during the $\beta$-decay of the metastable state can be found close to listed peaks (within the resolution of the detector). Accordingly, the derived results may overestimate the ground state half-life of $^{95}$Tc. Finally, the half-life of $^{95m}$Tc has unusually high uncertainty, exceeding 3\,\% \cite{uni59}. Therefore, a high precision experiment has been carried out aiming at the determination of the half-lives of the above listed two isotopes and the isomeric state of $^{95}$Tc. This paper is organized as follows. In Section 2 the details of the experiment are presented. Section 3 shows the data analysis and summary is given in Section 4 along with the obtained results.


\section{Experimental details}
\label{sec:exp}

\begin{figure*}
\resizebox{1.0\textwidth}{!}{\includegraphics{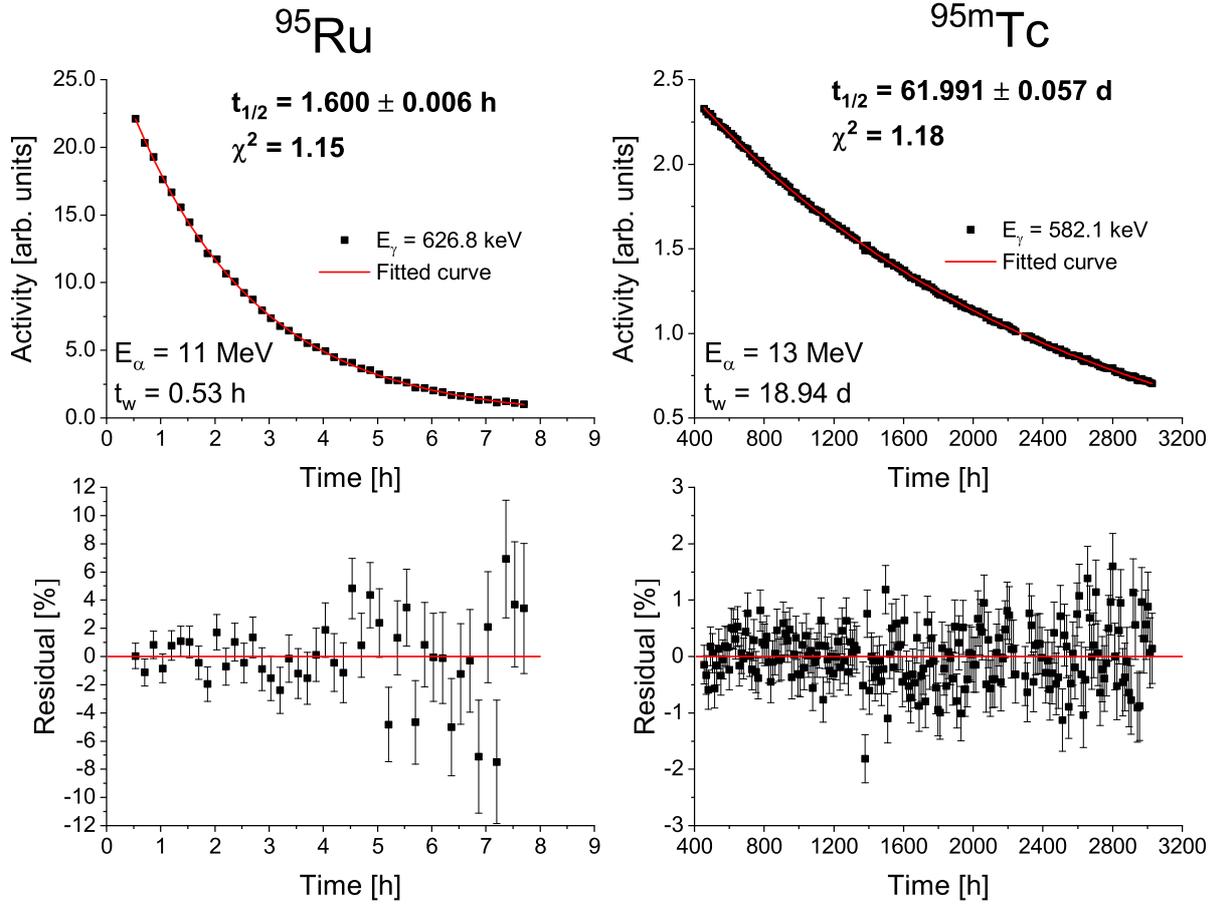}}
\caption{Decay curves (and residuals) based on the study of the E$_{\gamma}$\,=\,626.8\,keV ($^{95}$Ru) and E$_{\gamma}$\,=\,582.1\,keV ($^{95m}$Tc) $\gamma$-transitions. The widths of each time bin are 10 min for $^{95}$Ru and 12 hours for $^{95m}$Tc.}
\label{fig:decay1}       
\end{figure*}

The half-lives were derived using $\gamma$-spectroscopy. The $\gamma$-radiation following the $\beta$-decay of the reaction products was measured using a 50\,\% relative efficiency HPGe detector (Canberra model GR10024), placed into a complete 4$\pi$ multilayered shielding. After the irradiation, the samples were placed at 21\,cm distance from the endcap of the detector. The nuclear electronics used for the present half-life measurement was similar to the one used in the past for the precise half-life measurements carried out at Atomki \cite{gyu09,far11,gyu12,sze19}. Namely, the signals from the detector preamplifier were first shaped (the shaping time was set to 3$\mu$s) and amplified by an ORTEC model 671 module. An ORTEC model ASPEC-927 multichannel analyzer unit was used and the data were collected using the ORTEC A65-B32 MAESTRO software which provides deadtime information found to be precise in the previous works. 

\begin{table}
\caption{Decay parameters of the studied isotopes taken from \cite{NDS,NDS2,Kan10} ($^+$: emitted after the IT decay of the metastable state).}
\label{tab:decay}       
\begin{tabular}{llll}
\noalign{\smallskip}
Product & Decay mode &E$_{\gamma}$ [keV] & I$_{\gamma}$ [\%] \\
\noalign{\smallskip}\hline\noalign{\smallskip}
\hline
$^{95}$Ru	& 100\,\% $\beta$$^+$/$\epsilon$& 336.4	  & 69.9\,$\pm$\,0.5 \\
	        && 626.8	  & 17.8\,$\pm$\,0.5 \\
	        && 1096.8  & 20.9\,$\pm$\,1.0 \\				
\hline
$^{95}$Tc	&100\,\% $\beta$$^+$/$\epsilon$& 204.1   & 0.304\,$\pm$\,0.023 \\
	        && 765.8   & 93.8\,$\pm$\,0.3 \\
	        && 947.7   & 1.95\,$\pm$\,0.02 \\				
\hline
$^{95m}$Tc&96.12\,$\pm$\,0.32\,\% $\beta$$^+$/$\epsilon$ & 204.1	  & 63.2\,$\pm$\,0.8 \\
	        &3.88\,$\pm$\,0.32\,\% IT &582.1    & 30.0\,$\pm$\,0.4 \\
					&&616.5	  & 1.284\,$\pm$\,0.021 \\
	        &&765.8$^+$   & 93.8\,$\pm$\,0.3 \\
					&&786.2	  & 8.66\,$\pm$\,0.12 \\
	        &&820.6	  & 4.71\,$\pm$\,0.06 \\
	        &&1039.3	  & 2.78\,$\pm$\,0.04 \\
\hline				
$^{97}$Ru	&100\,\% $\beta$$^+$/$\epsilon$& 215.7 & 87.66\,$\pm$\,0.11 \\
	        && 324.5	& 11.01\,$\pm$\,0.12 \\				
\hline
$^{103}$Ru&100\,\% $\beta$$^-$& 497.1	& 91.0\,$\pm$\,1.2 \\
\hline
\noalign{\smallskip}
\end{tabular}
\vspace*{5cm}  
\end{table}

The product of the $^{92}$Mo($\alpha$,n) reaction is $^{95}$Ru, the $\beta$$^+$/$\epsilon$-decay of this isotopes leads to the ground and \linebreak metastable state of $^{95}$Tc. The $\beta$$^+$/$\epsilon$-decay of $^{95}$Tc results in $^{95}$Mo which is stable. The $^{95m}$Tc state decays by $\beta$$^+$/$\epsilon$-decay to $^{95}$Mo and with internal transition (IT) to the ground state of $^{95}$Tc, the branching ratios of these decay modes are I\,=\,96.12\,$\pm$\,0.32\,\% ($\beta$$^+$/$\epsilon$) and I\,=\,3.88\,$\pm$\,0.32\,\% (IT), respectively. Each decay is followed by the emission of several high relative intensity $\gamma$-rays. Moreover, $^{95}$Tc and $^{95m}$Tc can not only be produced or populated by the $\beta$-decay of $^{95}$Ru, but also directly during the irradiation of $^{92}$Mo through the $^{92}$Mo($\alpha$,p) reaction (although its cross section is about an order of magnitude lower than the cross section of the ($\alpha$,n) reaction). The relevant decay parameters of $^{95}$Ru, $^{95}$Tc and $^{95m}$Tc are summarized in Table \ref{tab:decay}. 

The $^{95}$Ru and $^{95}$Tc samples were produced by alpha irradiation of high purity natural isotopic composition 0.5 mm thick molybdenum plates. The energies, intensities and the length of the irradiations were optimized to produce samples with appropriate activities for the half-life measurements. On the one hand side, the activity of the sample should be high enough to measure the selected $\gamma$-transitions with good statistics. On the other hand, the deadtime of the counting system should be kept as low as possible to limit the systematic uncertainty arising from the deadtime correction. According to this, at first, using E$_{\alpha}$\,=\,12.0\,MeV and E$_{\alpha}$\,=\,11.0\,MeV $\alpha$ irradiations, relatively weak sources were produced. In the first 10 hours of the counting the spectra were stored and erased every 10 minutes (and from the yield of the E$_{\gamma}$\,=\,336.4\,keV, E$_{\gamma}$\,=\,626.8\,keV and E$_{\gamma}$\,=\,1096.8\,keV $\gamma$-transitions, the half-life of $^{95}$Ru was derived). Later the spectra were saved every hour. The spectra measured t$_w$\,=\,24 hours after the end of the irradiation were used to derive the half-life of $^{95}$Tc (during this waiting time practically all $^{95}$Ru isotopes already decayed to $^{95}$Tc).

Next, a high activity source was produced with a \linebreak E$_{\alpha}$\,=\,13.0\,MeV $\alpha$ beam irradiation and used to derive the half-life of $^{95m}$Tc. The source activity was measured for more than 3.5 months and the spectra were stored and erased every 12 hours. One day after the end of the irradiation the deadtime was still exceeding 20\,\%, therefore, this sample was solely used to derive the half-life of the metastable state and accordingly the activity measurement started about three weeks after the end of the irradiation. During this waiting time practically all $^{95}$Tc isotopes originated from the $\beta$-decay of $^{95}$Ru and from the $^{92}$Mo($\alpha$,p) reaction already decayed to $^{95}$Mo. Using the E$_{\gamma}$\,=\,204.1\,keV, E$_{\gamma}$\,=\,582.1\,keV, E$_{\gamma}$\,=\,616.5\,keV, \linebreak E$_{\gamma}$\,=\,786.2\,keV, E$_{\gamma}$\,=\,820.6\,keV and E$_{\gamma}$\,=\,1039.3\,keV and the E$_{\gamma}$\,=\,765.8\,keV (belonging to $^{95}$Tc) $\gamma$-transitions listed in Table \ref{tab:decay}, the half-life of the metastable state was derived with high precision. 
Figure \ref{fig:decay1} shows the decay curves of $^{95}$Ru and $^{95m}$Tc, the fitted exponential functions and the residuals.
 
\begin{figure}
\resizebox{0.5\textwidth}{!}{\includegraphics{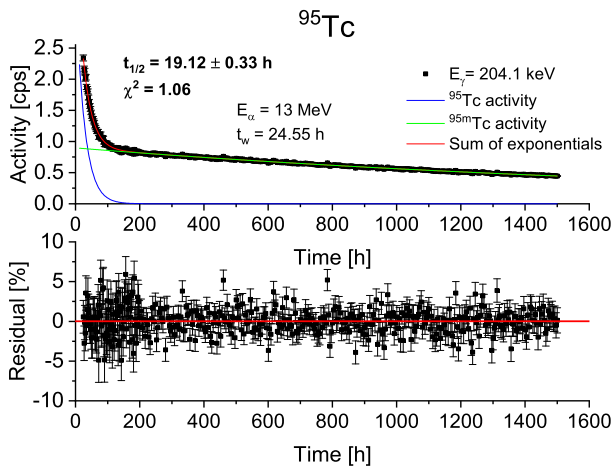}}
\caption{Decay curve (and residual) obtained from the measured E$_{\gamma}$\,=\,204.1\,keV $\gamma$-transition of $^{95}$Tc. The width of each time bin is 2 hours before and 4 hours after 188.00 hours long counting. }
\label{fig:decay2}       
\end{figure}

Finally, a moderate activity source was used to derive the ground state half-life of $^{95}$Tc. Although the $\gamma$-counting started about 34 min after the end of the irradiation, only the $\gamma$-spectra measured t$_w$\,=\,24.55 hours after the end of the irradiation were analyzed. During this waiting time, practically all produced $^{95}$Ru nuclei decayed to $^{95}$Tc. The total length of this counting was 62.17 days and to derive the half-life of $^{95}$Tc the yields of the E$_{\gamma}$ = 204.1 keV, E$_{\gamma}$\,=\,765.8\,keV and E$_{\gamma}$\,=\,947.7\,keV $\gamma$-transitions were measured. However, the $\beta$-decay of $^{95}$Ru populates not only the ground but a low-lying metastable state in $^{95}$Tc, too. Although the half-life of the metastable state is about 75 times longer than the one characterizing the ground state and furthermore the probability of the internal transition decay is rather low, still the $\gamma$-rays emitted along this decay chain influence the determination of the ground state half-life. According to this, the measured $\gamma$-yields (similarly to the E$_{\alpha}$\,=\,12.0\,MeV and E$_{\alpha}$\,=\,11.0\,MeV $\alpha$ beam irradiation) were fitted with the sum of two exponential functions. 
Figure \ref{fig:decay2} shows the decay curve of $^{95}$Tc and the sum of the two fitted exponential functions.

\begin{table}
\caption{Half-life results (HA refers to the E$_{\alpha}$\,=\,13.0\,MeV high activity source and MA refers for the E$_{\alpha}$\,=\,13.0\,MeV moderate activity source). The listed uncertainties are statistical only. For more details see text.}
\label{tab:results}       
\begin{tabular}{lllll}
\hline\noalign{\smallskip}
Isotope & E$_{\alpha}$ & E$_{\gamma}$ & t$_{1/2}$ & $\chi_{red.}^2$ \\
& [MeV] & [keV] & & \\
\noalign{\smallskip}\hline\noalign{\smallskip}
\hline
$^{95}$Ru&11.0	&	336.4	  &	1.6044\,$\pm$\,0.0018\,h		&	1.61	\\
         &      &	626.8	  &	1.6000\,$\pm$\,0.0055\,h		&	1.15	\\
         &	    &	1096.8	&	1.6074\,$\pm$\,0.0064\,h		&	0.84	\\
         &12.0	&	336.4	  &	1.6028\,$\pm$\,0.0032\,h		&	1.02	\\
         &      &	626.8	  &	1.5938\,$\pm$\,0.0087\,h		&	1.29	\\
         &			&	1096.8	&	1.5925\,$\pm$\,0.0094\,h		&	1.58	\\
\hline	
$^{95}$Tc	&	13.0 MA	&	204.12	&	19.12\,$\pm$\,0.33\,h		&	1.06	\\
	&		            &	765.8	  &	19.2579\,$\pm$\,0.0049\,h	&	1.04	\\
	&		            &	947.7	  &	19.280\,$\pm$\,0.056\,h 	&	1.08	\\
	&	12.0	        &	204.1	  &	18.84\,$\pm$\,0.77\,h	  &	1.02	\\
	&	11.0  	      &	204.1	  &	19.65\,$\pm$\,1.25\,h	  &	0.97	\\
\hline	
$^{95m}$Tc	&	13.0 HA	&	204.12	&	61.949\,$\pm$\,0.035\,d	&	1.18	\\
	&		&	582.1	&	61.991\,$\pm$\,0.057\,d		&	1.18	\\
	&		&	616.5	&	62.40\,$\pm$\,0.39\,d		&	1.16	\\
	&   & 765.8 & 62.05\,$\pm$\,0.18\,d    & 1.19  \\
	&		&	786.1	&	61.98\,$\pm$\,0.25\,d		&	1.02	\\
	&		&	820.6	&	61.99\,$\pm$\,0.16\,d		&	1.10	\\
	&		&	1039.3	&	61.82\,$\pm$\,0.21\,d		&	0.97	\\
	&	13.0 MA	&	204.1	&		61.69\,$\pm$\,0.24\,d	&	1.11\\
  &         & 765.8 &   62.5\,$\pm$\,1.1\,d    & 0.97 \\
\hline											
$^{97}$Ru&11.0	&	215.7	  &	2.8387\,$\pm$\,0.0028\,d	&	0.96	\\
         &12.0	&	215.7	  &	2.8422\,$\pm$\,0.0012\,d	&	1.37	\\
         &13.0	&	215.7	  &	2.8411\,$\pm$\,0.0004\,d	&	1.47	\\
\hline	
$^{103}$Ru&13.0	&	497.1	  &	39.297\,$\pm$\,0.021	d &	1.08	\\
\noalign{\smallskip}\hline
\end{tabular}
\vspace*{5cm}  
\end{table}
               

\section{Data analysis and results}
\label{sec:res}

\begin{figure*}
\resizebox{1.0\textwidth}{!}{\includegraphics{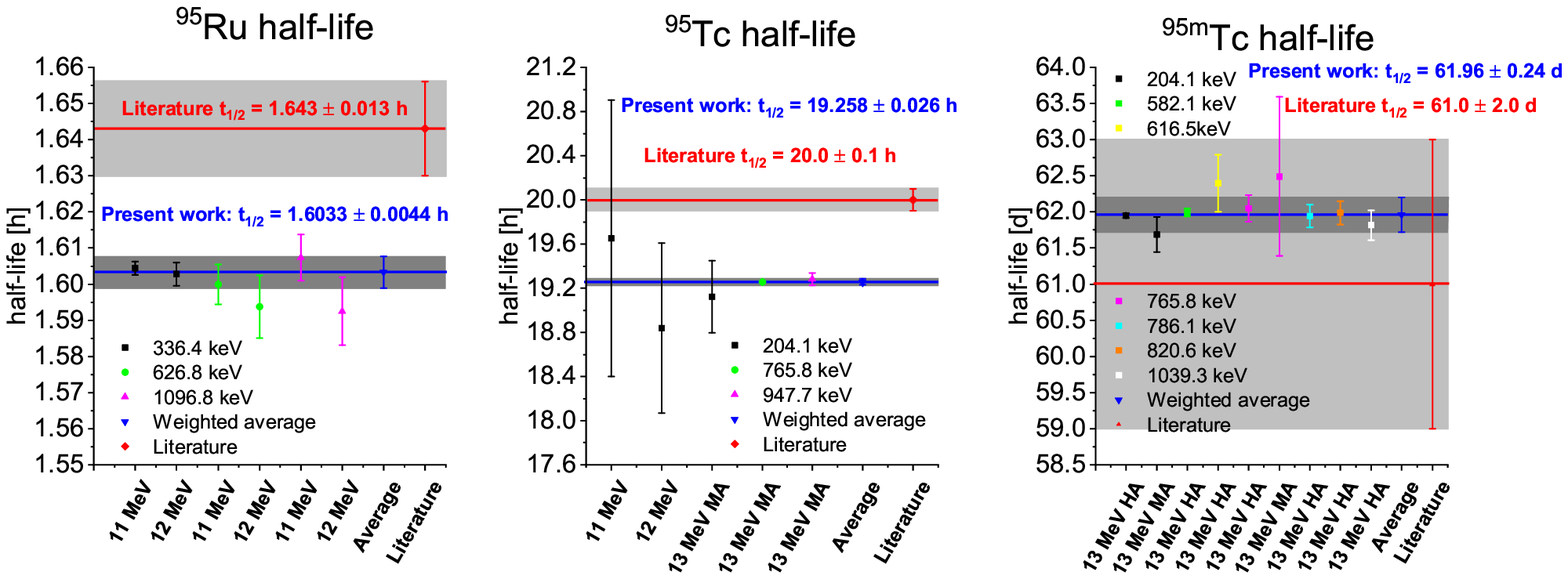}}
\caption{Obtained half-life values for $^{95}$Ru, $^{95}$Tc and $^{95m}$Tc. The plotted values are the individual results as well as the weighted averages of the half-lives determined from the analysis of the different $\gamma$-transitions listed in Table \ref{tab:decay}. The horizontal lines indicate the final half-life results, the grey shaded areas show the total uncertainties. For comparison, the literature half-life values and their uncertainties (in light grey) are also indicated.}
\label{fig:results}       
\end{figure*}

In each spectrum the peaks were fitted with a Gaussian plus a linear background. Moreover, numerical integration was also used to determine the areas. For both approaches different background regions were selected. The two methods resulted in half-lives always within 0.09\,\% which is taken into account as systematic uncertainty characterizing the peak area determination. To determine the half-lives of $^{95}$Ru and $^{95m}$Tc, exponential curves were fitted --- using the least square method (described in details e.g. in \cite{leo92}) --- to the deadtime corrected areas of the $\gamma$-peaks as the function of time. In the case of $^{95}$Tc, the measured $\gamma$-yields were fitted with the sum of two exponential functions, keeping the half-life of $^{95}$Tc and the initial number of the $^{95,95m}$Tc isotopes as free parameters (and the half-life of $^{95m}$Tc as fixed). Table \ref{tab:results} lists and Fig. \ref{fig:results} shows the measured half-life values for $^{95}$Ru, $^{95}$Tc and $^{95m}$Tc. The uncertainties, listed in Table \ref{tab:results} are statistical only and the $\chi^2_{red.}$ value refers to the exponential fit\footnote{In the calculation of the weighted mean, the uncertainties listed in table 2 have been scaled up by the square root of the corresponding $\chi_{red}$ for $\chi_{red}$ $>$ 1 cases.}. As can be seen the values derived from the different $\gamma$-transitions are in good agreement. The final t$_{1/2}$ values and their statistical uncertainties were calculated as the weighted average of the studied transitions.

Systematic uncertainties can influence the half-life results. For example any change in the detection efficiency during the $\gamma$-counting leads to the alteration of the results. By measuring the half-life of a reference source this possible effect and also the precision of the deadtime determination can be studied. Natural isotopic composition molybdenum targets were used in our experiment, thus, $^{97,103}$Ru isotopes were also produced via ($\alpha$,n) reactions on $^{94,100}$Mo. The half-lives of these ruthenium isotopes were measured recently with very high precision \cite{god09} and found to be t$_{1/2}$\,=\,2.8370\,$\pm$\,0.0014\,d (0.049\,\%)\footnote{Relative uncertainties are given in parenthesis.} and t$_{1/2}$\,=\,39.210 \linebreak \,$\pm$\,0.038\,d (0.097\,\%), respectively. Their $\beta$-decay is followed by the emission of high intensity $\gamma$-rays suitable for half-life determination. Our resulted t$_{1/2}$ values for the $^{97,103}$Ru reference isotopes are listed in Table \ref{tab:results}, too. As it can be seen, our data are higher than the literature values by 0.14\,\% and 0.22\,\%, respectively, while our statistical uncertainties are 0.015\,\% and 0.054\,\%. In order to quantify the above discussed effect, the deadtime values provided by the data acquisition system (DAQ) were modified and its effect on the obtained half-lives were studied. We found that the literature half-lives of the reference sources can be reached by modifying the deadtime by 18.3\% ($^{97}$Ru) and 23.3\% ($^{103}$Ru). The effect of such deadtime modifications on the half-lives of the studied isotopes are 0.24\%, 0.10\% and 0.38\% for $^{95}$Ru, $^{95}$Tc and $^{95m}$Tc, respectively. These values are taken as systematic uncertainties characterizing the stability of the counting setup and the deadtime determination.



Furthermore, any difference between the t$_{1/2}$ values based on the analysis of the different $\gamma$-transitions could be explained if e.g. we assume that weak $\gamma$-rays, emitted after the $\beta$-decay of other isotopes, with energies within the resolution of the HPGe detector contribute to yield of the peak of our interest (see e.g. \cite{sze19}). However, the derived half-life values, based on the countings of different $\gamma$-transitions are statistically consistent and this indicates that the measured $\gamma$-yields contain no parasitic contributions and therefore, systematical uncertainty is not assigned.


\section{Conclusions}
\label{sec:con} 

The half-lives of $^{95}$Ru, $^{95}$Tc and $^{95m}$Tc were measured by high precision using $\gamma$-spectroscopy. Table \ref{tab:results} lists and Fig. \ref{fig:results} shows the measured half-life values. The total uncertainties are the quadratic sum of the statistical uncertainties and the above discussed systematic uncertainties. The final half-life results are t$_{1/2}$\,=\,1.6033\,$\pm$\,0.0017\,(stat.)\,$\pm$\, \linebreak 0.0041\,(syst.)\,hours, t$_{1/2}$\,=\,19.258\,$\pm$\,0.005\,(stat.)\,$\pm$\,0.026\, \linebreak (syst.)\,hours, and t$_{1/2}$\,=\,61.96\,$\pm$\,0.03\,(stat.)\,$\pm$\,0.24\,(syst.)\, \linebreak days for $^{95}$Ru, $^{95}$Tc and $^{95m}$Tc, respectively. For the first two isotopes, the differences between the adopted values and our results exceed several $\sigma$ uncertainty and this fact calls for an independent confirmation. Our new $^{95m}$Tc half-life value is in agreement with the literature data, however, its uncertainty is lower by more than an order of magnitude, therefore, the result of this work is recommended as new adopted value. 

\section{Acknowledgments}

This work was supported by NKFIH (NN128072) and by the \'UNKP-19-4-DE-65, \'UNKP-19-3-I-DE-394 New National Excellence Program of the Ministry of Human Capacities of Hungary. G. G. Kiss acknowledges support from the J\'anos Bolyai research fellowship of the Hungarian \linebreak Academy of Sciences.

\end{document}